\documentstyle[aps,prl,multicol,epsf]{revtex}
\begin{document}
\draft

\newcommand{\pp}[1]{\phantom{#1}}
\newcommand{\be}{\begin{eqnarray}}
\newcommand{\ee}{\end{eqnarray}}
\newcommand{\ve}{\varepsilon}
\newcommand{\vs}{\varsigma}
\newcommand{\Tr}{{\,\rm Tr\,}}
\newcommand{\pol}{\frac{1}{2}}

\title{
Von Neumann equations with time-dependent Hamiltonians
and supersymmetric quantum mechanics
}
\author{Marek~Czachor$^{1,2}$, Heinz-Dietrich~Doebner$^{2}$, 
Monika~Syty$^{1,2}$, and Krzysztof~Wasylka$^{1,2}$}
\address{
$^1$ Wydzia{\l} Fizyki Technicznej i Matematyki Stosowanej\\
Politechnika Gda\'{n}ska,
ul. Narutowicza 11/12, 80-952 Gda\'{n}sk, Poland\\
$^2$ Arnold Sommerfeld Institut f\"ur Mathematische Physik\\
Technische Universit\"at Clausthal, 38678 Clausthal-Zellerfeld,
Germany
}
\maketitle

\begin{abstract}
Starting with a time-independent Hamiltonian $h$ and an appropriately
chosen solution of the  von Neumann equation  $i\dot\rho(t)=[ h,\rho(t)]$
we construct its binary-Darboux partner $h_1(t)$ and an exact 
scattering solution of  
$i\dot\rho_1(t)=[h_1(t),\rho_1(t)]$ where $h_1(t)$ is time-dependent and
not isospectral to $h$. 
The method is analogous to supersymmetric
quantum mechanics but is based on a different version of a Darboux 
transformation. We illustrate the technique by
the example where $h$ corresponds to a 1-D harmonic oscillator. 
The resulting $h_1(t)$ represents a 
scattering of a soliton-like pulse on a three-level system.
\end{abstract}


\begin{multicols}{2}
One of the main ideas of supersymmetric (SUSY) quantum mechanics (QM) can be 
summarized as follows \cite{SUSY}. 
Assume we know a ground state $|\psi_0\rangle$ of a
stationary Schr\"odinger equation (SE)
\be
H|\psi_0\rangle=
(H_{\rm kin}+V)|\psi_0\rangle=E_0|\psi_0\rangle \label{SE_1}
\ee
with some $V$ and $E_0$. Using $|\psi_0\rangle$ we construct 
an ``annihilation'' operator
$A=A(\psi_0)$ satisfying 
$
A|\psi_0\rangle=0
$
and $H-E_0=A^{\dag}A$. 
Now define $|\psi_1\rangle:=A|\psi\rangle$
(here $|\psi\rangle$ is any eigenstate of $H$ linearly independent of 
$|\psi_0\rangle$, with eigenvalue $E$) 
and  $H_1=AA^{\dag}=H_{\rm kin}+V_1$. $H_1$ is 
the so-called SUSY partner Hamiltonian of $H$. 
Then, using $AH=AA^{\dag}A=H_1A$, one finds that 
\be
(H_{\rm kin}+V_1)|\psi_1\rangle=E|\psi_1\rangle.\label{SE_2}
\ee
In a single step we have produced a new potential $V_1$ and 
one solution of the corresponding stationary SE. 

The map $V\to V_1$ is known to be particular example 
of a {\it Darboux transformation\/} (DT) 
\cite{MS}. All DT transform a ``potential'' $V$ into 
$V_1$ and simultaneously generate an ``annihilation'' operator $A(\psi_0)$ 
satisfying $A(\psi_0)\psi_0=0$, where $\psi_0$ is a solution of some 
partial differential linear equation associated with $V$. The 
physical interpretation of such an 
abstract ``potential'' depends on the problem. 

SUSY QM deals with {\it linear\/} SE and for this 
reason the density matrix generalization is not interesting: 
$H_1$ can be inserted either into the
SE or into the von Neumann equation (vNE) 
$
i\dot \rho=[H_1,\rho].\label{vNE1}
$

However, the vNE has a structure which is algebraically different from this
of the SE and therefore allows for different DT. 
A candidate is the so-called binary DT (BDT) originally constructed in 
\cite{LU} and applied to 
optical soliton equations. 
Quite recently the technique was applied to Yang-Mills equations 
\cite{U} and nonlinear vNE \cite{SLMC,MKMCSL}. 
A tutorial introduction to density matrix applications is 
given in \cite{MCMKSLJN}. There are formal 
analogies between the BDT and the ``dressing method'' of 
Zakharov {\it et al.\/}
\cite{Z} but technically the two procedures are inequivalent
(for a discussion cf. \cite{U,MKMCSL}).

The purpose of the Letter is to show that the
BDT leads to a new kind of 
SUSY-type QM for density matrices which does not have a counterpart 
in SUSY QM based on SE. At an intermediate stage 
of the construction we solve the nonlinear vNE
\be
i\dot \rho &=& [H,\rho^2],\label{EAvNE}
\ee
where $H$ is a time-independent Hamiltonian.  
The set of solutions 
of (\ref{EAvNE}) contains all the pure states of standard QM since for 
$\rho^2=\rho$ (\ref{EAvNE}) reduces to the linear vNE. For
$\rho^2\neq\rho$ there exist at least two more classes of solutions. 
One of them occurs for $\rho$'s satisfying either $\rho^2-a\rho=0$ 
with $a\in \bbox R$,
$a\neq 1$, or a weaker condition $[H,\rho^2-a\rho]=0$ 
(now $a=1$ is acceptable).
In both cases $\rho(t)=e^{-iaHt}\rho(0)e^{iaHt}$. The second class is 
of the form  $\rho(t)=e^{-iaHt}\rho_{\rm int}(t)e^{iaHt}$
with $\rho_{\rm int}(-\infty)\neq\rho_{\rm int}(+\infty)$. These
additional solutions, 
called here the self-scattering (SS) ones, are 
fundamental to our construction because of the following property: 
Each SS solution of the {\it nonlinear\/} vNE (\ref{EAvNE}) with the 
time-{\it independent\/} $H$ is simultaneously a scattering solution of a
{\it linear\/} vNE with a time-{\it dependent\/} Hamiltonian $h_1(t)$. 
Both the SS solution and the new Hamiltonian are algebraically constructed 
in terms of BDT. The construction does not make use of 
supercharges and for this reason the resulting partner 
Hamiltonians will be termed the binary-Darboux (BD) partners.

The BDT method of solving 
(\ref{EAvNE}) was described in \cite{SLMC,MKMCSL}. 
We start with the family of Lax pairs, parametrized by 
$\omega\in \bbox C$, 
\be
z_\omega |\psi_\omega\rangle 
&=& 
(\rho-\omega H)|\psi_\omega\rangle,\label{1a}\\
i|\dot\psi_\omega\rangle 
&=&
\Big(
H\rho +\rho H-\omega H^2\Big)|\psi_\omega\rangle,
\label{1b}
\ee
where $z_\omega$ is a complex eigenvalue. 
The pair (\ref{1a})-(\ref{1b}) is here the
analogue of (\ref{SE_1}); $\rho$ and $h=H\rho +\rho H$ play the role of the 
``potentials''. 

The connection of (\ref{1a})--(\ref{1b}) with (\ref{EAvNE}) is two-fold. 
First, the 
necessary condition for the existence of $|\psi_\omega\rangle$
is given by (\ref{EAvNE}). 
Now assume $|\psi_\mu\rangle=|\psi_\mu(t)\rangle$ is any solution 
of (\ref{1a})--(\ref{1b}) with 
$\omega=\mu$ and some $z_\mu$.
Denote by $P_\mu$ the projector on  
$|\psi_\mu\rangle$ and let   
$\lambda\in \bbox C$ be another parameter,
and $\rho$ any solution of (\ref{EAvNE}). 
Defining
\be
\rho_1 
&=&
\Big({\bbox 1}+\frac{\mu-\bar\mu}{\bar\mu}P_\mu\Big)
\rho
\Big({\bbox 1}+\frac{\bar\mu-\mu}{\mu}P_\mu\Big)
=:
U_\mu\rho U_\mu^{\dag}\label{rho_1}\\
|\psi_{\lambda,1}\rangle
&=&
\Big({\bbox 1}-\frac{\mu-\bar\mu}{\lambda-\bar\mu}P_\mu\Big)
|\psi_\lambda\rangle
=:
A(\psi_\mu)|\psi_\lambda\rangle\label{A}
\ee
we find (cf. \cite{SLMC,MKMCSL})
\be
z_\lambda |\psi_{\lambda,1}\rangle
&=& 
(\rho_1-\lambda H)|\psi_{\lambda,1}\rangle\label{1a'}\\
i|\dot\psi_{\lambda,1}\rangle
&=&
\Big(
H\rho_1 +\rho_1 H-\lambda H^2\Big)|\psi_{\lambda,1}\rangle.\label{1b'}
\ee
The ``Hamiltonians'' 
$\rho-\lambda H$ and $\rho_1-\lambda H$ possess 
the same eigenvalue $z_\lambda$ and their eigenvectors are related by
the ``annihilation operator'' $A$ (note that $A(\psi_\mu)|\psi_\mu\rangle=0$). 
However, these are not the physical BD partners we are interested 
in. BDT transforms  the two ``potentials'' $\rho\to\rho_1$, $h\to h_1$ 
in such a way that 
\be
i\dot \rho_1 &=&\big[H\rho_1+\rho_1 H,\rho_1\big]
=
\big[h_1,\rho_1\big],\label{EAvNE_1}
\ee
since this condition has to be satisfied whenever $|\psi_{\lambda,1}\rangle$
exists. The BD-transformed  Lax pair (\ref{1a'})-(\ref{1b'}) can be used to
repeat the procedure: $\rho_1\to\rho_2$, $h_1\to h_2$. 

To explicitly show that the construction of $h_1$ is nontrivial 
we have to make an assumption 
about the Hamiltonian $H$. We shall concentrate on the isospectral family 
of the 1-D harmonic oscillator (HO) since for Hamiltonians with 
equally-spaced spectrum  a strategy 
leading to nontrivial solutions was worked out in detail in \cite{SLMC}.
An alternative strategy was described in \cite{MKMCSL} and applied to a 
concrete example in \cite{MCMKSLJN}. 
In both cases the result is a SS solution. 

We take the Hamiltonian $H=\epsilon N$, where $\epsilon$ is some parameter,
\be
N=\sum_{n=0}^\infty (r+n)|r+n\rangle\langle r+n| \label{H}
\ee
and $r\in \bbox R$ (e.g. for 1-D HO $r=1/2$; for 3-D isotropic HO 
$r=3/2$). In the Hilbert space spanned by 
$\{|r+n\rangle\}_{n=0}^\infty$ consider a 3-D subspace 
spanned by three subsequent excited states 
$|k\rangle$, $|k+1\rangle$, and $|k+2\rangle$. It should be stressed that 
the same strategy can be applied to any $H$ with discrete spectrum provied 
there exist three eigenvalues of $H$ satisfying $E_k-E_l=E_l-E_m$.  

In order to obtain a SS solution $\rho_1(t)$ one has to start with an 
appropriate $\rho(t)$. The problem of how to select such a 
$\rho$  has been discussed in great detail in 
\cite{SLMC}. The fact that (\ref{solution}) does indeed solve 
(\ref{EAvNE}) with $H$ given by 
(\ref{H}) can be verified by a straightforward calculation. 

We consider a one-parameter family of solutions, parametrized by 
$\alpha\in \bbox R$. Physically the parameter turns out to control the 
scattering process. Mathematically it parametrizes an initial condition 
for the solution of the Lax pair (\ref{1a})-(\ref{1b}).
We solve (\ref{1a})-(\ref{1b}) with 
\be
\rho(t)&=&e^{-i5Ht}\rho(0)e^{i5Ht}\label{rho(t)}=:
W_5\rho(0)W_5^{\dag},\label{W_5}\\
\rho(0)&=&\frac{5}{2}\Big(|k\rangle\langle k|+
|k+2\rangle\langle k+2|\Big)\nonumber\\
&\pp =&
+
\frac{5+\sqrt{5}}{2}|k+1\rangle\langle k+1|\nonumber\\
&\pp =&
-\frac{3}{2}
\Big(|k+2\rangle\langle k|+|k\rangle\langle k+2|\Big),
\ee
and $\mu=i/\epsilon$. 
For later purposes we have introduced the unitary operator
$W_a(t)=e^{-iaHt}$. (\ref{rho(t)}) is a solution of (\ref{EAvNE}) 
and therefore
the necessary condition for the existence of $|\psi_i(t)\rangle$ is satisfied. 
$5H$ in (\ref{rho(t)}) comes from $[H,\rho(0)^2]=5[H,\rho(0)]$
and the resulting equalities
\be 
i\dot\rho=[H\rho+\rho H,\rho]=5[H,\rho]=[h,\rho].
\ee
$h=5H$ can be regarded as the first element of the pair of BD partner 
Hamiltonians we are going to find.
The initial condition for   (\ref{1a})-(\ref{1b}) is 
\be
|\psi_i(0)\rangle
&=&
\textstyle{\frac{1}{\sqrt{1+\alpha^2}}}
|k+1\rangle\nonumber\\
&\pp =&
+
\textstyle{\frac{\alpha}{\sqrt{1+\alpha^2}}}
\Big(
 -i\textstyle{\sqrt{\frac{3+\sqrt{5}}{6}}}
|k\rangle
+
\textstyle{\sqrt{\frac{2}{9+3\sqrt{5}}}}
|k+2\rangle\Big).\nonumber 
\ee
Inserting 
$P_i$, which projects  on $|\psi_i(t)\rangle$, into 
(\ref{rho_1}) with $\mu=i/\epsilon$ 
and normalizing the resulting solution to get 
$\Tr \rho_1=1$, we finally get the density matrix 
\be
\rho_1(t)=\sum_{u,v=0}^2\rho_1(t)_{1+u,1+v}
|k+u\rangle\langle k+v|\label{solution}
\ee
where the matrix of coefficients in (\ref{solution}) is
\be
\rho_1(t)_{\dots}
=
\frac{1}{15+\sqrt{5}}
\left(
\begin{array}{ccc}
5 & \xi(t) & \zeta(t) \\
\bar \xi(t) & 5+\sqrt{5} & \xi(t) \\
\bar \zeta(t) & \bar \xi(t) & 5
\end{array}
\right).\label{rhomatr}
\ee
with
\be 
\xi(t)
&=&
\frac{\left(2+3i-\sqrt{5}i\right)\sqrt{3+\sqrt{5}}\alpha}
{\sqrt{3}\big(e^{\omega_0 t/5}+\alpha^2
e^{-\omega_0 t/5}\big)}
e^{i\omega_0 t}
\label{xi}\\
\zeta(t)
&=&
-
\frac{9e^{2\omega_0 t/5}+\left(1+4\sqrt{5}i\right)\alpha^2}
{3\big(e^{2\omega_0 t/5}+\alpha^2\big)}
e^{2i\omega_0 t}.
\label{zeta}
\ee
and $\omega_0=10\epsilon/(15+\sqrt{5})$. Writing (\ref{solution}) as 
\be
\rho_1(t)&=&e^{-i\omega_0  Nt}
\rho_{\rm int}(t)
e^{i\omega_0 Nt}
\ee
one finds, for $0<|\alpha|<\infty$, 
\be
\rho_{\rm int}(-\infty)
&=&
\textstyle{
\frac{1}{15+\sqrt{5}}}
\left(
\begin{array}{ccc}
5 & 0 & -\frac{1}{3}-\frac{4\sqrt{5}i}{3} \\
0 & 5+\sqrt{5} & 0 \\
-\frac{1}{3}+\frac{4\sqrt{5}i}{3} & 0 & 5
\end{array}
\right),\nonumber\\
\rho_{\rm int}(+\infty)
&=&
\textstyle{
\frac{1}{15+\sqrt{5}}}
\left(
\begin{array}{ccc}
5 & 0 & -3 \\
0 & 5+\sqrt{5} & 0 \\
 -3 & 0 & 5
\end{array}
\right).\nonumber
\ee
This shows that (\ref{solution}) is a SS solution.

The BD partner $h_1$ occuring in (\ref{EAvNE_1}) is non-unique and defined  
up to an operator commuting with $\rho_1$. 
This freedom is useful. Set 
\be 
h_1=(H+\epsilon c_1\bbox 1)\rho_1+\rho_1(H+\epsilon c_1\bbox 1)
+\epsilon c_2\bbox 1
\ee
with constant $c_1$, $c_2$.  Denoting
$\sigma_{j,k}=|j\rangle\langle k|$, and using the
above explicit solution we find 
\be
h_1(t)
&=&
\tilde H
+
H_1(t)\label{h_1(t)}
\ee
where $\tilde H= \omega_0(N+c_1\bbox 1)
+\epsilon c_2\bbox 1$ and 
\be
H_1(t)
&=&
\frac{\omega_0}{\sqrt{5}}(k+c_1+1)\sigma_{k+1,k+1}
\nonumber\\
&\pp =&
+
\frac{\omega_0}{5}(k+c_1+\textstyle{\frac{1}{2}})\big(
\xi(t)
\sigma_{k,k+1}
+\bar \xi(t)
\sigma_{k+1,k}\big)\nonumber\\
&\pp =&
+\frac{\omega_0}{5}
(k+c_1+1)\big(\zeta(t)
\sigma_{k,k+2}+\bar\zeta(t)
\sigma_{k+2,k}\big)
\nonumber\\
&\pp =&
+\frac{\omega_0}{5}
(k+c_1+\textstyle{\frac{3}{2}})\big(\xi(t)
\sigma_{k+1,k+2}
+ \bar\xi(t)
\sigma_{k+2,k+1}\big).\nonumber
\ee
$\xi(t)$, $\zeta(t)$ are essentially 
the Rabi frequencies. The non-uniqueness of $h_1$ was used again to 
extend the nonperturbed part of (\ref{h_1(t)}) beyond the 3-D subspace.  
Our construction guarantees that (\ref{solution}) is a scattering 
solution of the 
corresponding time-dependent {\it linear\/} vNE
$
i\dot\rho_1=[h_1(t),\rho_1].
$ 
Let us note here that the dynamics of $\rho_1$ is related to $\rho(0)$ 
by the {\it unitary\/} transformation $U_iW_5$. 
In general, taking arbitrary $U_\mu$ and $W_a$ 
we can alternatively define the scattering 
Hamiltonian as
\be
h_1=i\dot U_\mu U_\mu^{\dag}+aU_\mu H U_\mu^{\dag}.
\ee
$h_1$ is a nontrivial scattering Hamiltonian provided  $\rho_1(t)$
is a SS solution of (\ref{EAvNE}).   

(\ref{h_1(t)}) represents a complicated time-dependent 
three-level perturbation of a HO. 
In order to better understand the kind of interaction we have produced
set $\alpha=1$, $c_1+k+1=0$, $\epsilon c_2=-\omega_0 c_1$, 
and  $d=\sqrt{3+\sqrt{5}}(2+3i-\sqrt{5}i)
/(2\sqrt{3})$. 
The Hamiltonian now reads
\be
h_1(t)
&=&
\omega_0 N
-
\frac{\omega_0de^{i\omega_0t}}{10\cosh(\omega_0t/5)}
\big(\sigma_{k,k+1}-\sigma_{k+1,k+2}\big)
\nonumber\\
&\pp =&
\pp{\omega_0 H}
-\frac{\omega_0d^*e^{-i\omega_0t}}{10\cosh(\omega_0t/5)}
\big(\sigma_{k+1,k}-\sigma_{k+2,k+1}\big).\label{sech}
\ee
One can think of $h_1$ as describing 
a 1-D HO located at $x=0$ 
and interacting with the well-known McCall-Hahn ``sech'' optical soliton
\cite{McCH}. Let us recall, however, that the result is more general and 
valid for any $H$ with 
discrete spectrum provided the 3-D subspace corresponds to three equally
spaced eigenvalues.
Taking different parameters in $h_1$ we obtain additional terms 
reminiscent of the ``sech-tanh'' pulse occuring in inhomogeneously 
broadened three-level media \cite{Eberly}. It is interesting that 
for $c_1+k+1\neq 0$ the perturbation $H_1(t)$ contains a time-independent 
term proportional to $|k+1\rangle\langle k+1|$. Redefining the non-perturbed 
part by 
\be
\tilde H'=\omega_0 N
+
\frac{\omega_0}{\sqrt{5}}(k+c_1+1)|k+1\rangle\langle k+1|
\ee
we break the equal spacing of the non-perturbed Hamiltonian, simultaneously 
detuning the highest and the lowest levels from $\omega_0$ and generating
a transition with doubled frequency $2\omega_0$.

A general property of (\ref{EAvNE}) is the fact that 
$\langle H\rangle_n =\Tr H \rho^n$ are integrals of motion for any 
natural $n$ and any solution $\rho$ \cite{MCJN}. 
In particular, this implies that the sum of the perturbed eigenvalues
of $h_1$  
is time-independent. The same holds for the average energy
$\langle E\rangle= \Tr h_1(t)\rho_1(t)$. However, the eigenvalues 
themselves  may be time-dependent. For $c_1+k+1=0$, $c_2=0$, 
the eigenvalues of the restriction of $h_1$ to the 3-D subspace 
are 0 and 
$$
\pm\frac{\omega_0}{5}\sqrt{25+4 \frac{e^{2\omega_0t/5}\alpha^2}
{(e^{2\omega_0t/5}+\alpha^2)^2}}.
$$
This implies that 
the BD partners $h=5H$ and $h_1$ are not isospectral, a situation that may 
occur in higher-dimensional SUSY. 

The figures illustrate properties of the scattering solutions.
Fig.~1 shows the average position of the 1-D HO
$\langle x\rangle=\frac{1}{\sqrt{2}}
\Tr\rho_1(a+a^{\dag})$ as a function of time and $\alpha$. 
In the asymptotic regions the average is 0. For times where 
$\langle x\rangle\approx 0$ 
the dynamics is effectively given by 
\be
\rho_{\rm in}(t)
&=&
e^{-i\omega_0Nt}\rho_{\rm int}(-\infty)e^{i\omega_0Nt}\nonumber\\
\rho_{\rm out}(t)
&=&
e^{-i\omega_0Nt}\rho_{\rm int}(+\infty)e^{i\omega_0Nt}\nonumber.
\ee
As $|\alpha|$ grows the moment of SS is shifted towards the 
future. For $\alpha=0$ or $|\alpha|=\infty$ there is no scattering
since $\rho_{\rm int}$ becomes time independent. 

The asymptotic probability 
densities in position 
space $\varrho(x,t)=\langle x|\rho_1(t)|x \rangle$ are symmetric (implying 
$\langle x\rangle=0$), Fig.~2. 
Such time-dependent probability distributions represent a new type of 
nonlinear effect. We propose to term them the Harzians \cite{Harzian}.

The above effects can be extended to higher-dimensional 
subspaces. One of the 
possibilities is related to the ``weak superposition'' principle: 
For any family of solutions $\{\rho_k\}$ of (\ref{EAvNE}) satisfying 
$\rho_k\rho_l=0$ for $k\neq l$, the combination 
$\rho(t)=\sum_k p_k\rho_k(p_k t)$ is also a solution of (\ref{EAvNE}). 
One can generalize the procedure to many noninteracting HO and  
consideration of systems with degeneracy, such as HO 
with spin, leads to a nontrivial second iteration of BDT: 
$\rho\to\rho_1\to\rho_2$ and $h\to h_1\to h_2$. 
Another possibility is related to the Yang-Mills (YM) case.
The result of \cite{U} shows that a class of YM equations can be integrated by 
BDT. The anti-self-dual YM case is algebraically related to Euler-Arnold
equations \cite{Mason} which are a particular case of (\ref{EAvNE}) 
as discussed in \cite{SLMC}. 

Exactly solvable equations with time dependent Hamiltonians are a rarity 
in quantum mechanics. 
The technique we have described leads to a broad class of 
such equations. The example we have discussed, in spite of 
its simplicity, shows the richness and efficiency of the method. 
The resulting three-level dynamics is highly nontrivial and physically 
interesting. We expect the method to prove useful in many branches of quantum
physics. 

The work of M.C. was financed by the Humboldt Foundation and is a part of 
the KBN project No. 2 P03B 163 15 and the Flemish-Polish joint project
No. 007. The work of M.S. and K.W. was made possible by the international 
student exchange program financed by the German Academic Exchange Service 
(DAAD). 
We thank Jan Naudts, Sergiej Leble, Maciek Kuna,  Partha Guha, Krishna 
Maddaly, and Ary Perez for discussions.

\narrowtext  

\begin{figure}
\epsfxsize=8.25cm
\epsffile{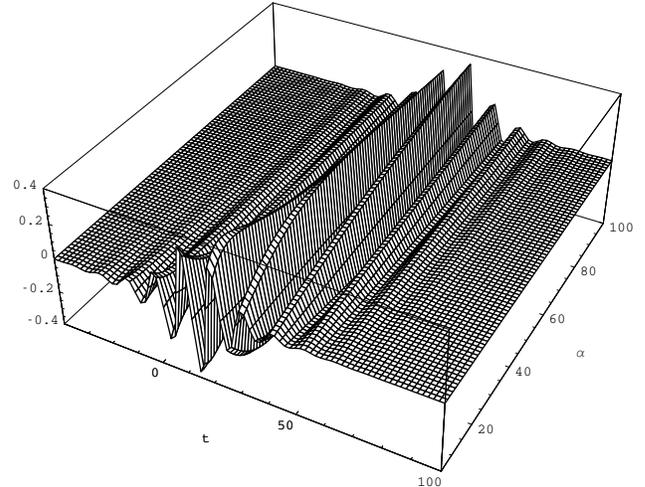}
\caption{$\langle x\rangle$  
as a function of time and the parameter 
$\alpha$, $5\leq \alpha\leq 100$, which
controls the initial condition. The moment of SS moves 
towards the future (past) as $|\alpha|$ grows (decreases). 
For $|\alpha|=\infty$ ($\alpha=0$)
SS is shifted to $+(-)\infty$ (no scattering).}
\end{figure}
\begin{figure}
\epsfxsize=8.25cm
\epsffile{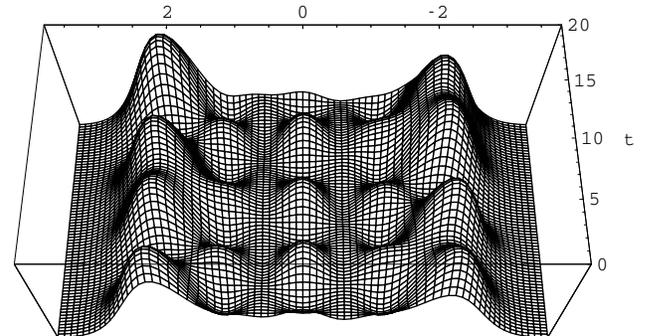}
\caption{The Harzian. Probability density in position space as a function of 
time for $k=2.5$, $\alpha=5$,
$0<t<20$. 
The asymmetry of the probability density is responsible for the oscillation of 
$\langle x\rangle$ seen at Fig.~1.}
\end{figure}
\begin{figure}
\epsfxsize=8.25cm
\epsffile{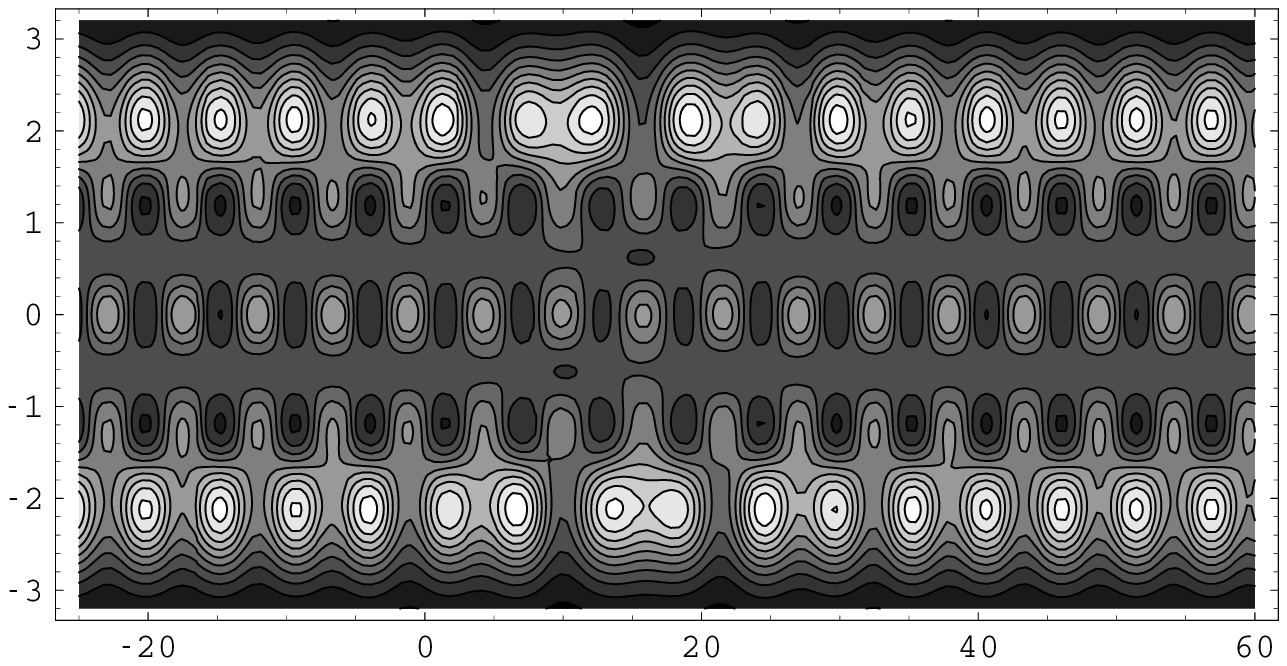}
\caption{Contour plot of the Harzian from Fig.~2 for $-25<t<60$. The 
continuous transition between the two asymptotic states (with symmetric 
probability distributions) is clearly visible.}
\end{figure}
\end{multicols}

\begin{references}
\bibitem{SUSY}F. Cooper, A. Khare, and U. Sukhatme, Phys. Rep. {\bf 251}, 
267 (1995), and references therein.
\bibitem{MS}V.~B.~Matveev, M.~A.~Salle, {\it Darboux
Transfomations and Solitons\/} (Springer, Berlin, 1991).
\bibitem{LU}S.~B.~Leble and N.~V.~Ustinov, in {\it
Nonlinear Theory and its Applications
(NOLTA~'93)\/}, p. 547 (Hawaii, 1993);
A.~A.~Zaitsev and S.~B.~Leble, Rep. Math. Phys. {\bf
39}, 177 (1997);
S.~B.~Leble, Computers Math. Applic. {\bf 35}, 73
(1998).
\bibitem{U}N. V. Ustinov, J. Math. Phys. {\bf 39}, 976 (1998).
\bibitem{SLMC}S.~B.~Leble and M.~Czachor, Phys. Rev. E {\bf 58}, 7091
(1998).
\bibitem{MKMCSL}M. Kuna, M. Czachor, and S.~B.~Leble, 
Phys. Lett. A {\bf 255}, 42 (1999).
\bibitem{MCMKSLJN}M. Czachor, M. Kuna, S.~B.~Leble, and J. Naudts, 
in {\it New Insights in Quantum Mechanics\/}, H.-D.~Doebner {\it et al.\/}
eds. (World Scientific, Singapore, 1999); quant-ph/9904110.
\bibitem{Z}S. P. Novikov, S. V. Manakov, L. P. Pitaevski, and V. E. Zakharov,
{\it Theory of Solitons, the Inverse Scattering Method\/} 
(Consultants Bureau, New York, 1984).
\bibitem{MCJN}M.~Czachor and J. Naudts, Phys. Rev. E {\bf 59}, 2497R 
(1999). 
\bibitem{McCH}S. L. McCall and E. L. Hahn, 
Phys. Rev. Lett. {\bf 18}, 908 (1967).
\bibitem{Eberly}A. Rahman and J. H. Eberly, Phys. Rev. A {\bf 58}, R805 
(1998).
\bibitem{Harzian}The mountain range shape from Fig.~2 sugested
an association with the 
Harz Mountains, where Arnold Sommerfeld Institute is located and this work 
was done.
\bibitem{Mason}L. J. Mason and N. M. J. Woodhouse, {\it Integrability, 
Self-Duality, and Twistor Theory\/} (Oxford, 1996). 
\end{references}
\end{document}